\documentclass[11pt]{article}%
\usepackage{amsmath}
\usepackage{amsfonts}
\usepackage{amssymb}
\usepackage{graphicx}%
\providecommand{\U}[1]{\protect\rule{.1in}{.1in}}

\begin{document}

\title{Bayesian Hypothesis Assessment in Two-arm Trials Using Relative Belief Ratios}
\author{Saman Muthukumarana and Michael Evans\thanks{ Saman Muthukumarana is Assistant
Professor, Department of Statistics, University of Manitoba, Winnipeg,
Manitoba R3T 2N2, Canada. Michael Evans is Professor, Department of
Statistics, University of Toronto, Toronto, Ontario, M5S 3G3, Canada. The
authors were partially supported by research grants from the Natural Sciences
and Engineering Research Council of Canada. }}
\date{}
\maketitle

\begin{center}
\textbf{Abstract}\bigskip
\end{center}

\noindent This paper develops a Bayesian approach for assessing equivalence
and non-inferiority hypotheses in two-arm trials using relative belief ratios.
A relative belief ratio is a measure of statistical evidence and can indicate
evidence either for or against a hypothesis. In addition to the relative
belief ratio, we also compute a measure of the strength of this evidence as a
calibration of the relative belief ratio. Furthermore, we make use of the
relative belief ratio as a measure of evidence, to assess whether a given
prior induces bias either for or against a hypothesis. Prior elicitation,
model checking and checking for prior-data conflict procedures are developed
to ensure that the choices of model and prior made are relevant to the
specific application. We highlight the applicability of the approach and
illustrate the proposed method by applying it to a data set obtained from a
two-arm clinical trial.\bigskip


\section{Introduction}

Recently, hypothesis testing has been an active research topic in various
types of two-arm clinical trials. As an example, a clinician may want to
demonstrate whether a new treatment is not worse than that of a reference
treatment (also known as active control or standard treatment) by more than a
specified margin [1]. This helps in assessing whether a less toxic, easier to
administer, or less expensive treatment, is medically noninferior to a
reference treatment. This kind of clinical trial, where the intention is to
demonstrate that the new treatment is not inferior to the standard treatment
by more than a small predefined margin $\delta,$ is known as a noninferiority
trial. Here $\delta>0$ is known as the prespecified clinically irrelevant or
noninferiority margin. Two-arm noninferiority trials of a new treatment and a
well established reference treatment are an attractive option as in certain
settings they avoid exposing patients to placebo situations. There has been a
series of articles on this topic; see for example, special issues of
Statistics in Medicine (Volume 47, Issue 1, 2005) and Journal of
Biopharmaceutical Statistics (Volume 14, Number 2, 2004).

Sometimes in clinical trials, the goal is not to show that the new treatment
is better, but rather equivalent to the reference treatment. This kind of
clinical trial is known as an equivalence trial [2]. In an equivalence trial,
the aim is to show that new treatment and the reference treatment have equal
efficacy. For practical purposes, one can select a margin $\delta$ such that
two treatments can be considered not to differ when their true difference lies
in the interval of clinical equivalence $(-\delta,\delta).$ Note that this is
different from testing a difference between two treatments which is a
two-sided test known as superiority test in clinical literature. In this case
$\delta>0$ is called the clinically equivalence margin. Note that in either
case, $\delta$ must be defined \textit{a priori}.

There is a considerable literature on the related problems of hypothesis tests
in clinical trials. As a simple frequentist method, one can use the standard
$t$ test for testing these hypotheses. At a higher level a generalized p-value
approach may be applied using a generalized test function [3]. One can also
perform tests using the ratio of the averages instead of the difference
between the averages [4, 5]. Bayesian non-inferiority tests for proportions in
two-arm trials with a binary primary endpoint [6] and normal means [7] are
also considered in recent literature.

This paper considers a novel Bayesian approach for assessing a hypothesis in
two-arm trials using relative belief ratios. A relative belief ratio for a
hypothesized value of a parameter of interest is interpreted as the evidence
that the hypothesized value is correct. We may obtain evidence either for or
against the hypothesized value. Associated with a relative belief ratio is a
measure of the strength of this evidence and this may be weak, strong or
inconclusive. General inferences based on relative belief ratios are developed
in [8, 9, 10].

We discuss relative belief ratios and associated theory in Section 2 and
illustrate this by application to the problem of interest. In Section 3 we
consider the elicitation of the prior and how we can measure the suitability
of the prior both with respect to its agreement with what the data say and
with respect to measuring the bias a prior puts into the analysis. In Section
4, the approach is applied to a data set obtained from a two-arm clinical
trial. We conclude with a short discussion in Section 5.

\section{Inferences Based on Relative Belief Ratios}

Suppose we have a statistical model, as given by a collection of densities
$\{f_{\theta}:\theta\in\Theta\},$ and a prior $\pi$ on $\Theta.$ After
observing data $x,$ the posterior distribution of $\theta$ is given by the
density
\[
\pi(\theta\,|\,x)=\frac{\pi(\theta)f_{\theta}(x)}{m(x)}%
\]
where $m(x)=\int_{\Theta}\pi(\theta)f_{\theta}(x)\,d\theta.$ For an arbitrary
parameter of interest $\psi=\Psi(\theta)$ we denote the prior and posterior
densities of $\psi$ by $\pi_{\Psi}$ and $\pi_{\Psi}(\cdot\,|\,x),$
respectively. The relative belief ratio for a hypothesized value $\psi_{0}$ of
$\psi$ is defined by
\begin{equation}
RB_{\Psi}(\psi_{0})=\frac{\pi_{\Psi}(\psi_{0}\,|\,x)}{\pi_{\Psi}(\psi_{0})},
\label{relbel}%
\end{equation}
the ratio of the posterior to the prior at $\psi_{0}.$ As such, $RB_{\Psi
}(\psi_{0})$ is measuring how beliefs have changed that $\psi_{0}$ is the true
value from \textit{a priori} to \textit{a posteriori}. Considering the case
when the prior for $\psi$ is discrete, we have that $RB_{\Psi}(\psi_{0})>1$
means that the data have lead to an increase in the probability that $\psi
_{0}$ is correct, and so we have evidence in favor of $\psi_{0},$ while
$RB_{\Psi}(\psi_{0})<1$ means that the data have lead to a decrease in the
probability that $\psi_{0}$ is correct, and so we have evidence against
$\psi_{0}.$ As discussed in [10], this interpretation is also appropriate in
the continuous case via a consideration of limits.

Clearly relative belief ratios are similar to Bayes factors, as they are both
measuring change in belief, but the Bayes factor does this by comparing
posterior to prior odds while the relative belief ratio compares posterior to
prior probabilities and so is somewhat simpler. In fact, in certain
circumstances the relative belief ratio and Bayes factor can be considered as
equivalent but this is not always true. The full relationship between these
quantities is discussed in [10].

One problem that both the relative belief ratio and the Bayes factor share as
measures of evidence, is that it is not clear how they should be calibrated.
Certainly the bigger $RB_{\Psi}(\psi_{0})$ is than 1, the more evidence we
have in favor of $\psi_{0}$ while the smaller $RB_{\Psi}(\psi_{0})$ is than 1,
the more evidence we have against $\psi_{0}.$ But what exactly does a value of
$RB_{\Psi}(\psi_{0})=20$ mean? It would appear to be strong evidence in favor
of $\psi_{0}$ because beliefs have increased by a factor of 20 after seeing
the data. But what if other values of $\psi$ had even larger increases? While
calibrations of Bayes factors have been suggested [11, 12, 13] the proposed
scales seem arbitrary and it is not at all clear that there is a universal
scale on which Bayes factors or relative belief ratios can be calibrated.

A\ more useful calibration of (\ref{relbel}) is given by%
\begin{equation}
\Pi_{\Psi}(RB_{\Psi}(\psi)\leq RB_{\Psi}(\psi_{0})\,|\,x) \label{strength}%
\end{equation}
which is the posterior probability that the true value of $\psi$ has a
relative belief ratio no greater than that of the hypothesized value $\psi
_{0}.$ If we interpret $RB_{\Psi}(\psi_{0})$ as the measure of the evidence
that $\psi_{0}$ is the true value, we see that (\ref{strength}) is the
posterior probability that the true value has evidence no greater than that
for $\psi_{0}.$

While (\ref{strength}) may look like a p-value, we see that it has a very
different interpretation. For when $RB_{\Psi}(\psi_{0})<1,$ so we have
evidence against $\psi_{0},$ then a small value for (\ref{strength}) indicates
a large posterior probability that the true value has a relative belief ratio
greater than $RB_{\Psi}(\psi_{0})$ and so we have strong evidence against
$\psi_{0}.$ If $RB_{\Psi}(\psi_{0})>1,$ so we have evidence in favor of
$\psi_{0},$ then a large value for (\ref{strength}) indicates a small
posterior probability that the true value has a relative belief ratio greater
than $RB_{\Psi}(\psi_{0})$ and so we have strong evidence in favor of
$\psi_{0}.$ Notice that in the set $\{\psi:RB_{\Psi}(\psi)\leq RB_{\Psi}%
(\psi_{0})\},$ the `best' estimate of the true value is given by $\psi_{0}$
simply because the evidence for this value is the largest in this set. Various
results have been established in [10] supporting both (\ref{relbel}), as the
measure of the evidence and (\ref{strength}), as the strength of that evidence.

As a measure of the strength of the evidence, (\ref{strength}) seems to work
best when the posterior probabilities for all the possible values of $\psi$
are all small or even 0 as in the continuous case. When some of these values
have large posterior probabilities we can augment (\ref{strength}) as follows.
If the prior $\pi_{\Psi}$ corresponds to a discrete distribution with
$\pi_{\Psi}(\psi_{0})>0,$ we have that
\begin{equation}
\pi_{\Psi}(\psi_{0}\,|\,x)\leq\Pi_{\Psi}(RB_{\Psi}(\psi)\leq RB_{\Psi}%
(\psi_{0})\,|\,x)\leq RB_{\Psi}(\psi_{0}). \label{strength0}%
\end{equation}
The right-hand inequality holds generally, see [10], while the left-hand
inequality requires discreteness. Suppose $\pi_{\Psi}(\psi_{0}\,|\,x)$ and
(\ref{strength}) are both small and notice that this happens whenever
$RB_{\Psi}(\psi_{0})$ is small. In this case we clearly have strong evidence
against $\psi_{0}$ when $RB_{\Psi}(\psi_{0})<1$ and weak evidence for
$\psi_{0}$ when $RB_{\Psi}(\psi_{0})>1.$ Also, when $\pi_{\Psi}(\psi
_{0}\,|\,x)$ and (\ref{strength}) are both big, then we have only weak
evidence against $\psi_{0}$ when $RB_{\Psi}(\psi_{0})<1$ and strong evidence
for $\psi_{0}$ when $RB_{\Psi}(\psi_{0})>1.$

The other possibility is that the posterior probability $\pi_{\Psi}(\psi
_{0}\,|\,x)$ is small and (\ref{strength}) is big. If the prior probability
$\pi_{\Psi}(\psi_{0})$ is big and $RB_{\Psi}(\psi_{0})<1,$ then this suggests
that we have indeed obtained strong evidence against $\psi_{0}$ because
(\ref{strength}) is big only because there are many other values of $\psi$ for
which there is evidence against $\psi$ at least as strong as the evidence
against $\psi_{0}.$ If, however $\pi_{\Psi}(\psi_{0})$ is small and $RB_{\Psi
}(\psi_{0})<1,$ then we have weak evidence against $\psi_{0}$ because
$\pi_{\Psi}(\psi_{0}\,|\,x)$ is small due to $\pi_{\Psi}(\psi_{0})$ being
small. When $\pi_{\Psi}(\psi_{0})$ is big and $RB_{\Psi}(\psi_{0})>1,$ then we
must have $\pi_{\Psi}(\psi_{0}\,|\,x)$ is big as well, so no ambiguity arises,
while when $\pi_{\Psi}(\psi_{0})$ is small, then again $\pi_{\Psi}(\psi
_{0}\,|\,x)$ is small due to $\pi_{\Psi}(\psi_{0})$ being small and so in both
situations we have strong evidence in favor of $\psi_{0}$ via (\ref{strength}%
). So the only context where (\ref{strength}) might not suffice as a measure
of the strength of the evidence given by $RB_{\Psi}(\psi_{0}),$ is when $\psi$
has a discrete prior distribution with $\pi_{\Psi}(\psi_{0})$ a non-neglible
size, $\pi_{\Psi}(\psi_{0}\,|\,x)$ small and (\ref{strength}) big. In general
there is no harm in the discrete case in quoting (\ref{strength}), $\pi_{\Psi
}(\psi_{0}\,|\,x)$ and $\pi_{\Psi}(\psi_{0}),$ as part of the analysis of the
strength of the evidence given by $RB_{\Psi}(\psi_{0})$ and we recommend this.

There is another issue associated with using $RB_{\Psi}(\psi_{0})$ to assess
the evidence that $\psi_{0}$ is the true value. One of the key concerns with
Bayesian inference methods is that the choice of the prior can bias the
analysis in various ways. \ For example, in many problems Bayes factors and
relative belief ratios can be made arbitrarily large by choosing the prior to
be increasingly diffuse. This phenomenon is known as the Jeffreys-Lindley
paradox because a diffuse prior is supposed to represent less information.

An approach to dealing with this paradox is discussed in [10]. Given that we
accept that $RB_{\Psi}(\psi_{0})$ is the evidence that $\psi_{0}$ is true, the
solution is to measure \textit{a priori} whether or not the chosen prior
induces bias either for or against $\psi_{0}.$ To see how to do this we note
first the Savage-Dickey result, see [14] and [10], which says that
\begin{equation}
RB_{\Psi}(\psi_{0})=\frac{m(x\,|\,\psi_{0})}{m(x)} \label{savagedickey}%
\end{equation}
where
\[
m(x\,|\,\psi_{0})=\int_{\{\theta:\Psi(\theta)=\psi_{0}\}}\pi(\theta
\,|\,\psi_{0})f_{\theta}(x)\,d\theta
\]
is the prior-predictive density of the data $x$ given that $\Psi(\theta
)=\psi_{0}.$ Actually, it is easy to see that, if $T(x)$ is a minimal
sufficient statistic for the full model, then $m(x\,|\,\psi_{0})/m(x)=m_{T}%
(T(x)\,|\,\psi_{0})/m_{T}(T(x))$ where $m_{T}$ is the prior predictive density
of $T$ and $m_{T}(\cdot\,|\,\psi_{0})$ is the prior predictive density of $T$
given that $\Psi(\theta)=\psi_{0}.$

From (\ref{savagedickey}) we can measure the bias in the evidence against
$\psi_{0}$ by computing%
\begin{equation}
M_{T}\left(  \left.  \frac{m_{T}(t\,|\,\psi_{0})}{m_{T}(t)}<1\,\right\vert
\,\psi_{0}\right)  \label{bias1}%
\end{equation}
as this is the prior probability that we will obtain evidence against
$\psi_{0}$ when $\psi_{0}$ is true. So when (\ref{bias1}) is large we have
bias against $\psi_{0}.$ To measure the bias in favor of $\psi_{0}$ we choose
values $\psi_{0}^{\prime}\neq\psi_{0}$ such that the difference between
$\psi_{0}$ and $\psi_{0}^{\prime}$ represents the smallest difference of
practical importance. We then compute
\begin{equation}
M_{T}\left(  \left.  \frac{m_{T}(t\,|\,\psi_{0})}{m_{T}(t)}>1\,\right\vert
\,\psi_{0}^{\prime}\right)  \label{bias2}%
\end{equation}
as this is the prior probability that we will obtain evidence in favor of
$\psi_{0}$ when $\psi_{0}$ is false. Again, when (\ref{bias2}) is large we
have bias in favor of $\psi_{0}.$ Note that both (\ref{bias1}) and
(\ref{bias2}) decrease with sample size and so, in design situations, they can
be used to set sample size and so control bias.

When we are not able to control sample size, then (\ref{bias1}) and
(\ref{bias2}) can be computed and used to qualify any conclusions we reach
about whether $\psi_{0}$ is true or not. For example, if we have evidence
against $\psi_{0}$ and (\ref{bias1}) is large, this has to be taken with a
`grain of salt' as our choices have biased things this way. We draw a similar
conclusion if we have evidence in favor of $\psi_{0}$ and (\ref{bias2}) is
large. Of course, these negative conclusions could also lead us to redo the
analysis using a prior that does not induce such biases when this is possible,
see Section 3.

A variety of other inferences can be derived from interpreting $RB_{\Psi}%
(\psi)$ as the evidence that $\psi$ is the true value. For example, the best
estimate of $\psi$ is clearly the value for which the evidence is greatest,
namely,
\[
\psi_{LRSE}(x)=\arg\sup RB_{\Psi}(\psi),
\]
and called the least relative surprise estimator in [8, 9, 10]. Associated
with this is a $\gamma$-credible region
\begin{equation}
C_{\gamma}(x)=\{\psi:RB_{\Psi}(\psi)\geq c_{\gamma}(x)\} \label{credregion}%
\end{equation}
where
\[
c_{\gamma}(x)=\inf\{k:\Pi_{\Psi}(RB_{\Psi}(\psi)\geq k\,|\,x)\leq\gamma\}.
\]
Notice that $\psi_{LRSE}(x)\in C_{\gamma}(x)$ for every $\gamma\in\lbrack0,1]$
and so, for selected $\gamma,$ we can take the size of $C_{\gamma}(x)$ as a
measure of the accuracy of the estimate $\psi_{LRSE}(x).$ Given the
interpretation of $RB_{\Psi}(\psi)$ as the evidence for $\psi,$ we are forced
to use the sets $C_{\gamma}(x)$ for our credible regions. For if $\psi_{1}$ is
in such a region and $RB_{\Psi}(\psi_{2})\geq RB_{\Psi}(\psi_{1}),$ then we
must put $\psi_{2}$ into the region as well as we have at least as much
evidence for $\psi_{2}$ as for $\psi_{1}.$

In [8, 9, 10] various optimality properties are established for $\psi
_{LRSE}(x)$ and the regions $C_{\gamma}(x)$ in the class of all Bayesian
inferences. One notable property is that inferences based on the relative
belief ratio are invariant under reparameterizations. This is not the case for
Bayesian inferences based on losses, such as the posterior mean or mode and
highest probability density regions.

We now consider the application of relative belief inferences to two-arm
trials.\medskip\ 

\noindent\textbf{Example} \textit{Two-arm Trials.}

Let $x_{E}=(x_{E,1},\ldots,x_{E,n_{E}})$ denote the sample from the
experimental treatment and $x_{R}=(x_{R,1},\ldots,x_{R,n_{R}})$ denote the
sample from the reference treatment. We assume that these responses are
mutually independent with $x_{E,i}\sim N(\mu_{E},\sigma^{2})$ and $x_{R,i}\sim
N(\mu_{R},\sigma^{2})$ where $\mu_{E},\mu_{R}\in R^{1}$ and $\sigma^{2}>0$ are
all unknown. The information in the data is summarized by the minimal
sufficient statistic $T(x_{E},x_{R})=(\bar{x}_{E},\bar{x}_{R},s^{2})$ where
$s^{2}=[(n_{E}-1)s_{E}^{2}+(n_{R}-1)s_{R}^{2}]/(n_{E}+n_{R}-2)$ and the
likelihood equals%
\[
\sigma^{-n_{E}-n_{R}}\exp\{-\left[  n_{E}(\bar{x}_{E}-\mu_{E})^{2}+n_{R}%
(\bar{x}_{R}-\mu_{R})^{2}+(n_{E}+n_{R}-2)s^{2}\right]  /2\sigma^{2}\}.
\]
We will use the prior for $(\mu_{E},\mu_{R},\sigma^{2})$ given by%
\begin{align}
\mu_{E}\,|\,\sigma^{2}  &  \sim N(\mu_{0},\tau_{0}^{2}\sigma^{2}),\nonumber\\
\mu_{R}\,|\,\sigma^{2}  &  \sim N(\mu_{0},\tau_{0}^{2}\sigma^{2}),\nonumber\\
1/\sigma^{2}  &  \sim\text{Gamma}(\alpha_{0},\beta_{0}). \label{prior}%
\end{align}
We will discuss elicitation of the hyperparameters $\mu_{0},\tau_{0}%
^{2},\alpha_{0}$ and $\beta_{0}$ in Section 3. The posterior distribution of
$(\mu_{E},\mu_{R},\sigma^{2})$ is then easily obtained and is given
by\noindent%
\begin{align}
&  \mu_{E}\,|\,x_{E},x_{R},\sigma^{2}\sim N\left(  \frac{n_{E}\bar{x}_{E}%
+\mu_{0}/\tau_{0}^{2}}{n_{E}+1/\tau_{0}^{2}},\frac{\sigma^{2}}{n_{E}%
+1/\tau_{0}^{2}}\right)  ,\nonumber\\
&  \mu_{R}\,|\,x_{E},x_{R},\sigma^{2}\sim N\left(  \frac{n_{R}\bar{x}_{R}%
+\mu_{0}/\tau_{0}^{2}}{n_{R}+1/\tau_{0}^{2}},\frac{\sigma^{2}}{n_{R}%
+1/\tau_{0}^{2}}\right)  ,\nonumber\\
&  1/\sigma^{2}|\,x_{E},x_{R}\sim\text{Gamma}\left(  \frac{n_{E}+n_{R}%
+2\alpha_{0}}{2},\frac{2\beta_{0}+(n_{E}+n_{R}-2)s^{2}}{2}\right)  .
\label{posterior}%
\end{align}
Note that it is simple to generate values from (\ref{prior}) and
(\ref{posterior}).

Now suppose we want to assess the hypothesis that the true value of $\mu
_{E}-\mu_{R}$ satisfies ${|\mu_{E}-\mu_{R}|<\delta.}$ So ${\delta}$ represents
a practically meaningful difference between the means. If the difference is
less than this quantity, then we do not distinguish between $\mu_{E}$ and
$\mu_{R}$ but otherwise we do. It makes sense then that, if $\mu_{E}$ and
$\mu_{R}$ do differ, we would want to know how many units of ${\delta}$ these
means differed by. So for the $\psi$ parameter of interest in this problem we
will consider $\psi\in%
\mathbb{Z}
$ where $\psi=i$ indicates that ${\mu_{E}-\mu_{R}\in((2i-1)\delta
,(2i+1)\delta].}$ So the hypothesis of interest corresponds to $H_{0}:\psi=0.
$

To calculate the relative belief ratios for values of $\psi$ we need the prior
and posterior distributions of this parameter. These quantities are obtained
by discretizing the prior and posterior distributions of ${\mu_{E}-\mu_{R}.}$
We have that the marginal prior distribution of ${\mu_{E}-\mu_{R}}$ is given
by
\begin{equation}
({\mu_{E}-\mu_{R})/}\tau_{0}\sqrt{\frac{\beta_{0}}{\alpha_{0}}}\sim
t_{2\alpha_{0}} \label{priorpsi}%
\end{equation}
and the marginal posterior distribution of ${\mu_{E}-\mu_{R}}$ is given by
\begin{equation}
\{({\mu_{E}-\mu_{R})}-(\bar{x}_{E}-\bar{x}_{R})\}/s_{p}\sqrt{1/n_{E}+1/n_{R}%
}\,|\,x_{E},x_{R}\sim t_{\nu} \label{postpsi}%
\end{equation}
where $\nu=n_{E}+n_{R}+2\alpha_{0}-4$ and
\[
s_{p}^{2}=\frac{2\beta_{0}+(n_{E}+n_{R}-2)s^{2}}{n_{E}+n_{R}+2\alpha_{0}-4}.
\]
When $\nu$ is large, the posterior distribution is approximately normal, while
it has heavy tails when $\nu$ is small.

So to assess $H_{0}$ the evidence is given by%
\begin{equation}
RB_{\Psi}(0)=\frac{\Pi((-{\delta,\delta]}\,|\,x_{E},x_{R})}{\Pi((-{\delta
,\delta]})}=\frac{\Pi((-{\delta,\delta]}\,|\,\bar{x}_{E},\bar{x}_{R},s^{2}%
)}{\Pi((-{\delta,\delta]})} \label{relbel1}%
\end{equation}
and the strength of the evidence is given by%
\begin{align}
&  \Pi_{\Psi}(RB_{\Psi}(\psi)\leq RB_{\Psi}(\psi_{0})\,|\,x_{E},x_{R}%
)\nonumber\\
&  =\Pi(\cup_{RB_{\Psi}(i)\leq RB_{\Psi}(0)}{((2i-1)\delta,(2i+1)\delta
]}\,|\,x_{E},x_{R})\nonumber\\
&  =\sum_{i:RB_{\Psi}(i)\leq RB_{\Psi}(0)}\Pi(\{{((2i-1)\delta,(2i+1)\delta
]}\,|\,x_{E},x_{R}). \label{strength1}%
\end{align}
Both (\ref{relbel1}) and (\ref{strength1}) are easily evaluated using the
exact distribution theory given for the prior and posterior distribution of
${\mu_{E}-\mu_{R}.}$ For example, we can use the $t$ distribution function
routine in the R software package.

Suppose that we obtain $RB_{\Psi}(0)<1$ and that (\ref{strength1}) indicates
that this is reasonably strong evidence against $H_{0}.$ From the tabulation
of $RB_{\Psi}(i),$ that we computed as part of calculating (\ref{strength1}),
we easily obtain the optimal estimate of $\psi,$ namely,
\[
\psi_{LRSE}(x_{E},x_{R})=\arg\sup RB(i).
\]
If $\psi_{LRSE}(x_{E},x_{R})$ is greater than 0, then we have a clear
indication that the experimental treatment is better than the reference
treatment. The accuracy of the estimate is assessed by computing the
$0.95$-relative belief region
\[
C_{0.95}(x_{E},x_{R})=\{i:RB(i)\geq c_{0.95}(x_{E},x_{R})\}
\]
and seeing how large it is. We can convert this into a region for ${\mu
_{E}-\mu_{R}}$ via
\[
C_{0.95}^{\ast}(x_{E},x_{R})=\bigcup_{i\in C_{0.95}(x_{E},x_{R})}%
{((2i-1)\delta,(2i+1)\delta]}.
\]

To assess the bias in the prior we have to compute (\ref{bias1}) and
(\ref{bias2}). From (\ref{priorpsi}) and (\ref{postpsi}) we can evaluate
\[
RB_{\Psi}(0)=\frac{\Pi((-{\delta,\delta]}\,|\,x_{E},x_{R})}{\Pi((-{\delta
,\delta]})}=\frac{m_{T}(\bar{x}_{E},\bar{x}_{R},s^{2}\,|\,-{\delta<\mu_{E}%
-\mu_{R}\leq\delta})}{m_{T}(\bar{x}_{E},\bar{x}_{R},s^{2})}%
\]
and note that $RB_{\Psi}(0)$ depends on the data only through $(\bar{x}%
_{E}-\bar{x}_{R},s^{2}).$\textbf{\ }We then need only simulate from the
conditional prior predictive of $(\bar{x}_{E}-\bar{x}_{R},s^{2})$ given that
$\psi$ is the true value. Note that, given $(\mu_{E}-\mu_{R},\sigma^{2})$
then
\begin{align}
&  \bar{x}_{E}-\bar{x}_{R}\sim N(\mu_{E}-\mu_{R},(1/n_{E}+1/n_{R})\sigma
^{2}),\nonumber\\
&  (n_{E}+n_{R}-2)s^{2}/\sigma^{2}\sim\,\text{Chi-squared}(n_{E}+n_{R}-2)
\label{condpriorpred}%
\end{align}
and these quantities are independent.

We can compute (\ref{bias1}) by the following simulation process:

\begin{enumerate}
\item set a counter $C=0,$

\item generate $\sigma^{2}$ using (\ref{prior}),

\item generate $\mu_{E}-\mu_{R}$ from\thinspace a $N(0,2\tau_{0}^{2}\sigma
^{2})$ distribution conditioned to \newline$-{\delta<\mu_{E}-\mu_{R}\leq
\delta,}$

\item generate $(\bar{x}_{E}-\bar{x}_{R},s^{2})$ using (\ref{condpriorpred}),

\item compute $RB_{\Psi}(0)$ and add 1 to $C$ if it is less than 1,

\item repeat 2-4 $N$ times and record $C/N$ as the estimate of (\ref{bias1}).
\end{enumerate}

\noindent Essentially the same simulation can be carried out to evaluate
(\ref{bias2}) with step 2 changing, as we condition on ${(2i-1)\delta<\mu
_{E}-\mu_{R}\leq(2i+1)\delta}$ for say $i=1$ or $i=-1{,}$ and in step 4 we
check if $RB_{\Psi}(0)$ is greater than 1.

The only slightly difficult part in this simulation is step 3 and for that we
can use an inversion algorithm. For denoting the cdf and inverse cdf of a
$N(0,1)$ distribution by $\Phi$ and $\Phi^{-1},$ respectively, we generate
$\mu_{E}-\mu_{R}$ in step 3, when conditioning on ${(2i-1)\delta<\mu_{E}%
-\mu_{R}\leq(2i+1)\delta,}$ by generating $u\sim U(0,1)$ and putting
\begin{equation}
\mu_{E}-\mu_{R}=\Phi^{-1}(\Phi({(2i-1)\delta)+[}\Phi({(2i+1)\delta)-}%
\Phi({(2i-1)\delta)]u).} \label{invcdf}%
\end{equation}
We can use routines in $R$ for $\Phi$ and $\Phi^{-1}$ to evaluate
(\ref{invcdf}).

\section{Choosing and Checking the Ingredients}

In any statistical analysis a statistician chooses a model that supposedly
describes the generation of the data and, in a Bayesian analysis, also chooses
a prior. As the analysis is typically highly dependent on these subjective
choices, it is important that they be checked against what is typically
objective, at least if it is collected correctly, namely, the data.

\subsection{Checking the Model}

For the model this entails asking if the observed data is surprising for every
distribution in the model. If this is the case, then we conclude that there is
a problem with the model and need to somehow modify this. While there are
often many model checking procedures available, for the problem under study we
will use the Shapiro-Wilks test based on the residuals from the model. We note
that this check is, as it should be, completely independent of the choice of
prior as we do not want to confound our considerations of the adequacy of the
model and the prior.

\subsection{Eliciting the Prior}

Before discussing how we check the prior, we first consider the choice of the
prior. For this we need only consider eliciting the prior for $\mu$ and
$\sigma^{2}$ in a $N(\mu,\sigma^{2})$ distribution. So we need to specify the
hyperparameters $\mu_{0},\tau_{0}^{2},\alpha_{0}$ and $\beta_{0}.$ This is
based on knowledge of the measurement process that leads to the actual data
and will typically require knowledge from someone familiar with making these
kinds of measurements.

To elicit the prior for $\mu$ we specify an interval $(m_{1},m_{2})$ that we
are virtually certain (probability $=0.999$) will contain this quantity. Of
course we choose this as short as possible without being unrealistic. We then
set $\mu_{0}=(m_{1}+m_{2})/2$ and since
\[
0.999=\Phi\left(  \frac{m_{2}-\mu_{0}}{\tau_{0}\sigma}\right)  -\Phi\left(
\frac{m_{1}-\mu_{0}}{\tau_{0}\sigma}\right)  =2\Phi\left(  \frac{m_{2}-m_{1}%
}{2\tau_{0}\sigma}\right)  -1
\]
we have that
\begin{equation}
\sigma^{2}\leq\left(  (m_{2}-m_{1})/2\right)  ^{2}\left\{  \Phi^{-1}\left(
(1+0.999)/2\right)  \right\}  ^{-2}\tau_{0}^{-2}. \label{el1}%
\end{equation}
An interval that contains virtually all of the actual data measurements is
given by $\mu\pm\sigma\Phi^{-1}\left(  (1+0.999)/2\right)  .$ Since this
interval cannot be unrealistically too short or too long, we let $s_{1}$ and
$s_{2}$ be lower and upper bounds on the half-length of the interval so that
\begin{equation}
s_{1}^{2}\leq\sigma^{2}\left\{  \Phi^{-1}\left(  (1+0.999)/2\right)  \right\}
^{2}\leq s_{2}^{2}. \label{el2}%
\end{equation}
Then equating the upper bound on $\sigma^{2}$ in (\ref{el1}) with the upper
bound on $\sigma^{2}$ obtained from (\ref{el2}), we have
\[
\tau_{0}^{2}=\left(  (m_{2}-m_{1})/2\right)  ^{2}s_{2}^{-2}.
\]
This determines the hyperparameters in the conditional prior for $\mu.$

From (\ref{el2}) we have that%
\begin{equation}
s_{2}^{-2}\left\{  \Phi^{-1}\left(  (1+0.999)/2\right)  \right\}  ^{2}%
\leq1/\sigma^{2}\leq s_{1}^{-2}\left\{  \Phi^{-1}\left(  (1+0.999)/2\right)
\right\}  ^{2}. \label{el3}%
\end{equation}
Suppose again we want to determine the lower and upper bounds in (\ref{el3})
so that this interval contains $1/\sigma^{2}$ with virtual certainty.
Therefore, letting $G(\alpha_{0},\beta_{0},\cdot)$ denote the Gamma$(\alpha
_{0},\beta_{0})$ cdf we see that
\begin{align}
&  G^{-1}(\alpha_{0},\beta_{0},(1+0.999)/2)=s_{1}^{-2}\left\{  \Phi
^{-1}\left(  (1+0.999)/2\right)  \right\}  ^{2}\nonumber\\
&  G^{-1}(\alpha_{0},\beta_{0},(1-0.999)/2)=s_{2}^{-2}\left\{  \Phi
^{-1}\left(  (1+0.999)/2\right)  \right\}  ^{2} \label{el4}%
\end{align}
and we solve (\ref{el4}) for $\alpha_{0}$ and $\beta_{0}$\ by iteration.
Noting that $G(\alpha_{0},\beta_{0},x)=G(\alpha_{0},1,\beta_{0}x),$ and using
the monotonicity of the cdf, leads to a simple iterative process.

So specifying the hyperparameters for (\ref{prior}) requires specifying an
interval $(m_{1},m_{2})$ that contains the true values of $\mu_{E}$ and
$\mu_{R}$ with virtual certainty and also specifying the constants
$s_{1},s_{2}$ that specify lower and upper bounds on the length of any
interval that will contain any measurement with virtual certainty. Of course,
virtually certainty need not mean with probability $0.999$ as some other large
probability can be chosen. This value could be viewed as a conservative choice.

\subsection{Checking the Prior}

Checking the prior involves asking if the true value is a surprising value
with respect to the prior. Methods for checking the prior in this sense are
developed in [15, 17, 18]. Note that this is quite different than checking
whether or not a prior induces bias as discussed in Section 2. A prior can
avoid conflict with the data by being diffuse but at the same time induce bias
into the analysis. Selecting a suitable prior involves balancing these
considerations and tools have been developed for this.

The methods developed for checking the prior allows for all aspects of the
prior to be checked simultaneously or for checking separate aspects of the
prior in sequential fashion. The latter typically makes the most sense
because, if we do detect prior-data conflict, then we will be better able to
pinpoint where the problem lies.

The basic method for checking the prior involves computing, where $T$ is the
minimal sufficient statistic, the probability
\begin{equation}
M_{T}(m_{T}(t)\leq m_{T}(T(x))) \label{priorconflict}%
\end{equation}
as this serves to locate the observed value $T(x)$ in its prior distribution.
If (\ref{priorconflict}) is small, then $T(x)$ lies in a region of low prior
probability, such as a tail or anti-mode, which indicates a conflict. In the
continuous case (\ref{priorconflict}) is not invariant under 1-1 smooth
transformations of the minimal sufficient statistic. Accordingly,
(\ref{priorconflict}) was modified in [16] to
\begin{equation}
M_{T}\left(  m_{T}(t)/J_{T}(T^{-1}(t))\leq m_{T}(T(x))/J_{T}(x\right)  ),
\label{priorconflictmod}%
\end{equation}
where $J_{T}(x)=(\det dT(x)(dT(x))^{t})^{-1/2}$ and $dT(x)$ is the Jacobian
matrix of $T,$ to produce an invariant measure of prior-data conflict.

It is shown in [18] that, under quite general conditions, both
(\ref{priorconflict}) and (\ref{priorconflictmod}) converge to $\Pi(\pi
(\theta)\leq\pi(\theta_{true})),$ as the amount of data increases, where
$\theta_{true}$ is the true value of the parameter. If $\Pi(\pi(\theta)\leq
\pi(\theta_{true}))$ is small, then $\theta_{true}$ lies in a region of low
prior probability which implies that the prior is not appropriate. A logical
approach to modifying a prior to avoid a conflict, when this is detected, is
developed in [17].

For the prior given by (\ref{prior}) we will check this sequentially. First we
will check the prior on $\sigma^{2}$ and, if no prior-data conflict is found,
we then proceed to check the joint prior on $(\mu_{E},\mu_{R}).$ For this we
follow [15] which\ prescribes that the check on the prior for $\sigma^{2}$ is
based on the prior predictive distribution of $s^{2}.$ Given $\sigma^{2},$ we
have that $V=(n_{E}+n_{R}-2)s^{2}/\sigma^{2}\sim\,$Chi-squared$(n_{E}%
+n_{R}-2)$ and, as developed in the Appendix, a simple calculation gives that
the prior predictive distribution is $V\sim((n_{E}+n_{R}-2)/\alpha_{0}%
)\beta_{0}F((n_{E}+n_{R}-2)/2,2\alpha_{0}).$ Letting $m_{V}$ denote the
density of $V$ and following [16, 18] an invariant p-value that checks the
prior for $\sigma^{2}$ is given by
\begin{equation}
M_{V}(m_{V}(v)v^{1/2}\leq m_{V}((n_{E}+n_{R}-2)s^{2})(n_{E}+n_{R}-2)^{1/2}s)
\label{priorchk1}%
\end{equation}
and this is easily computed via simulation.

If the prior for $\sigma^{2}$ has passed its check, then we can proceed to
check the joint prior for $(\mu_{E},\mu_{R})$ and this is based on the prior
predictive distribution of $U=(\bar{x}_{E},\bar{x}_{R}).$ Given $\sigma^{2}$
we have that
\[
U\sim N_{2}\left(  \mu_{0}\left(
\begin{array}
[c]{c}%
1\\
1
\end{array}
\right)  ,\sigma^{2}\left(
\begin{array}
[c]{cc}%
\tau_{0}^{2}+1/n_{E} & 0\\
0 & \tau_{0}^{2}+1/n_{R}%
\end{array}
\right)  \right)
\]
and an easy calculation presented in the Appendix gives that the prior
predictive distribution is
\[
U\sim t_{2\alpha_{0}}\left(  2,\mu_{0}\left(
\begin{array}
[c]{c}%
1\\
1
\end{array}
\right)  ,(\beta_{0}/\alpha_{0})\left(
\begin{array}
[c]{cc}%
\tau_{0}^{2}+1/n_{E} & 0\\
0 & \tau_{0}^{2}+1/n_{R}%
\end{array}
\right)  \right)  .
\]
Denoting the prior predictive density of $U$ by $m_{U},$ we check the joint
prior for $(\mu_{E},\mu_{R})$ via the p-value
\begin{equation}
M_{U}(m_{U}(u)\leq m_{U}(\bar{x}_{E},\bar{x}_{R})) \label{priorchk2}%
\end{equation}
and this is easily computed using simulation. As discussed in [16, 18] this
p-value is invariant because $(\bar{x}_{E},\bar{x}_{R})$ is a linear function
of the data.

\section{Example}

We illustrate the approach described in Sections 2 and 3 using a data set
published in [2]. The data come from a comparative trial of moxonodin and
captopril in the antihypertensive treatment of patients suffering from major
depression. The response variable is the reduction of diastolic blood
pressure, measured in millimeters of mercury (mm Hg), of patients suffering
from a major depression under two drugs captopril (experimental) and moxonodin
(reference). The data is given by%
\begin{align*}
x_{E}  &  =(3.3,17.7,6.7,11.1,-5.8,6.9,5.8,3.0,6.0,3.5,18.7,9.6)\\
x_{R}  &  =(10.3,11.3,2.0,-6.1,6.2,6.8,3.7,-3.3,-3.6,-3.5,13.7,12.6)
\end{align*}
so $n_{E}=n_{R}=12.$ The minimal sufficient statistic of the data is
$T(x_{E},x_{R})=(\bar{x}_{E},\bar{x}_{R},s^{2})=(7.21,4.17,46.79).$ We suppose
that a practically meaningful difference in the means is given by $\delta=0.5$
mm Hg.

Figure 1 gives the Q-Q plots of the residuals. The Shapiro-Wilks test for
normality applied to the residuals gives a p-value of $0.51.$ This indicates
that the data is not inconsistent with the normality assumption with constant
variance.%
\begin{figure}
[ptb]
\begin{center}
\includegraphics[
height=2.3669in,
width=2.3669in
]%
{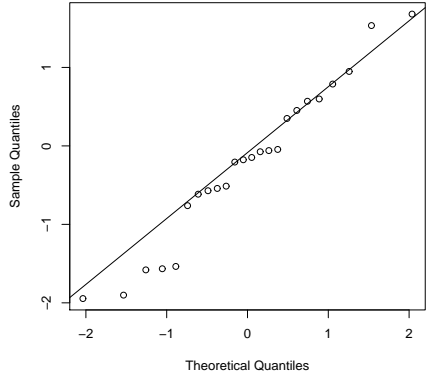}%
\caption{Quantile plots of experimental and reference treatment groups.}%
\end{center}
\end{figure}

For prior elicitation, we initially propose to reflect vague knowledge about
the parameters by choosing a very diffuse prior. The values $m_{1}%
=-100,m_{2}=100,s_{1}^{2}=5$ and $s_{2}^{2}=1000$ lead to the hyperparameter
values
\[
\mu_{0}=0,\tau_{0}^{2}=10,\alpha_{0}\approx2,\beta_{0}\approx5.
\]
The bias in this prior is assessed by computing (4) and (5) with $\psi_{0}=0$.
We found that
\[
M_{T}\left(  \left.  m_{T}(t\,|\,0)/m_{T}(t)<1\,\right\vert \,\psi=0\right)
=0.07,
\]
indicating that there is very little bias against $\psi_{0}.$ We see, however,
that
\[
M_{T}\left(  \left.  m_{T}(t\,|\,0)/m_{T}(t)>1\,\right\vert \,\psi=1\right)
=0.774
\]
and this indicates that we have considerable bias in favor of $\psi_{0}$. This
is undoubtedly because we have chosen the prior to be too diffuse. Using such
a prior will lead us to overstate the evidence in favor of the hypothesis or
equivalently, understate the evidence against.

So to avoid the bias as much as possible, we chose different values for the
hyperparameters. For this we set $m_{1}=-20,m_{2}=20,s_{1}^{2}=10$ and
$s_{2}^{2}=600$ which lead to the hyperparameter values
\[
\mu_{0}=0,\tau_{0}^{2}=0.67,\alpha_{0}\approx1,\beta_{0}\approx8.
\]
In this case we get
\begin{align*}
M_{T}\left(  \left.  m_{T}(t\,|\,0)/m_{T}(t)<1\,\right\vert \,\psi=0\right)
&  =0.49,\\
M_{T}\left(  \left.  m_{T}(t\,|\,0)/m_{T}(t)>1\,\right\vert \,\psi=1\right)
&  =0.40
\end{align*}
indicating that there is some bias both for and against the hypothesis with
this choice of prior. We cannot expect to be able to get both of these values
to be low as this is controlled by sample size and we do not have a lot of data.

We checked this prior using (\ref{priorchk1}) and (\ref{priorchk2}). The value
$0.15$ was obtained for (\ref{priorchk1}) and the value $0.22$ obtained for
(\ref{priorchk2}). This indicates that there is no reason to be concerned
about prior-data conflict.

Figure 2 contains a plot of the posterior and relative belief ratio for the
continuous parameter. For the discretized parameter $\psi$ we obtained
$RB_{\Psi}(0)=0.515$ with a strength of $0.19.$ As such we have only
moderately strong evidence against the hypothesis of equivalence. Also
\[
\psi_{LRSE}(x_{E},x_{R})=\arg\sup RB_{\Psi}(\psi)=7
\]
and the $0.95$-relative belief region is given by
\[
C_{0.95}^{\ast}(x_{E},x_{R})=(-0.5,13.5].
\]
The length of this interval indicates a fair degree of uncertainty about the
true value but we do have reasonable evidence that the treatments are not
equivalent.%
\begin{figure}
[ptb]
\begin{center}
\includegraphics[
height=2.5537in,
width=2.5537in
]%
{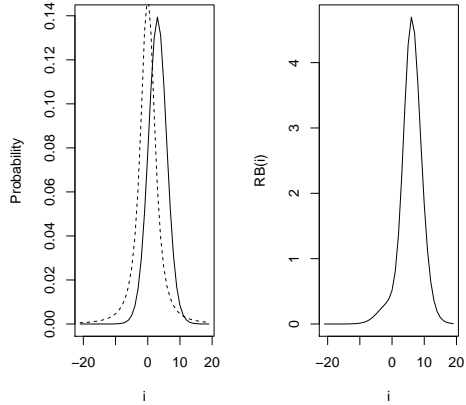}%
\caption{Plots of prior (- - -) and posterior (---) densities and of the
relative belief ratio (---) of $\psi$.}%
\end{center}
\end{figure}

Note that $C_{0.95}^{\ast}(x_{E},x_{R})$ includes the value 0 but this is not
a contradiction with the fact that we have evidence against $\psi=0.$ These
credible regions do not work like confidence intervals do with p-values. It is
only the length of $C_{0.95}^{\ast}(x_{E},x_{R})$ that is relevant as a
measure of the accuracy of the estimate $\psi_{LRSE}(x_{E},x_{R}).$ It is
easily deduced from (\ref{credregion}), and as discussed in [10], that there
is a relationship between relative belief regions and the strength of the
evidence (\ref{strength}). This makes sense as both are measuring the accuracy
or reliability of inferences based on the measure of evidence as given by the
relative belief ratio.

We investigated the choice of many other priors. The results were fairly
consistent in obtaining evidence against $\psi=0$ and with similar strengths.
The biases did vary considerably with sometimes there being very low bias
against $\psi=0$ but accompanied by high bias in favor or conversely.
Interestingly, such extreme cases are often accompanied by prior-data conflict.

Alternatively, we could assess the hypothesis of noninferiority, namely,
$H_{0}:\psi\in(-{\delta,\infty)}$ for which $\Pi((-{\delta,\infty)})=0.58$ and
$\Pi((-{\delta,\infty)}\,|\,x_{E},x_{R})=0.89.$ Therefore,%

\[
RB_{\Psi}((-{\delta,\infty)})=\frac{\Pi((-{\delta,\infty)}\,|\,x_{E},x_{R}%
)}{\Pi((-{\delta,\infty)})}=\frac{0.89}{0.58}=1.53
\]
indicates evidence in favor of $H_{0}$ being true. Since the posterior
probability of $H_{0}$ is large, this indicates strong evidence in favor of
the experimental treatment being at least as effective as the reference treatment.

\section{Conclusions}

We have considered the application of relative belief inferences to an
important inference problem with two-arm clinical trials. This is seen to
provide a clear definition of what the evidence is, whether in favor of or
against, for a hypothesis. Moreover, with such a definition this allows us to
assess the bias introduced into a statistical analysis by a proper prior and
so addresses a key concern with the use of Bayesian inference methods. In
addition we have provided a methodology for eliciting an appropriate prior in
such a context and demonstrated how one checks this prior to see if it is
contradicted by the data.

In general, we take the view that all of statistics is subjective as we choose
sampling models and priors and possibly even choose other ingredients. Such
subjectivity is always present in a statistical analysis. Rather than
searching for methods that are supposedly `objective' in some sense, we
embrace the subjectivity as allowing us to make judgements that reflect
additional information we have about the application. Once the model and prior
are chosen, we can go forward and make inference, based on the measure of
statistical evidence and its calibration, in an unambiguous way. If we want
these inferences to be convincing, however, it is important that we check the
ingredients chosen against that aspect of the analysis that can best be
claimed to be objective, namely, the data. So model checking and checking for
prior-data conflict are essential parts of any statistical analysis. The
analysis in this paper is presented as a meaningful application of this
approach to statistical analyses.

\section{Appendix}

In Section 3.3 the prior predictive density of $V=(n_{E}+n_{R}-2)s^{2}$ is%
\begin{align*}
m_{V}(v)  &  =\int_{0}^{\infty}\frac{(v/\sigma^{2})^{(n_{E}+n_{R}-2)/2-1}%
\exp\{-v/2\sigma^{2}\}}{2^{(n_{E}+n_{R}-2)/2}\Gamma((n_{E}+n_{R}-2)/2)}%
\frac{(1/\sigma^{2})^{\alpha_{0}}}{\beta_{0}^{\alpha_{0}}\Gamma(\alpha_{0}%
)}\times\\
&  \hspace{1in}\exp\{-\beta_{0}/\sigma^{2}\}\,d(1/\sigma^{2})\\
&  =\frac{\Gamma((n_{E}+n_{R}-2+2\alpha_{0})/2)}{\Gamma(\alpha_{0}%
)\Gamma((n_{E}+n_{R}-2)/2)}\left(  \frac{v}{2\beta_{0}}\right)  ^{(n_{E}%
+n_{R}-2)/2-1}\times\\
&  \hspace{1in}\left(  1+\frac{v}{2\beta_{0}}\right)  ^{-(n_{E}+n_{R}%
-2)/2-\alpha_{0}}\frac{1}{2\beta_{0}}.
\end{align*}

Suppose $U=(U_{1},U_{2})\,|\,\sigma^{2}\sim N_{2}(\mu,\sigma^{2}\Sigma)$ where
$\sigma^{2}$ is distributed as in (\ref{prior}). Then the marginal density of
$U$ is
\begin{align*}
m_{U}(u)  &  =\int_{0}^{\infty}(2\pi)^{-1}(\det\Sigma)^{-1/2}\exp
\{-(u-\mu)^{\prime}\Sigma^{-1}(u-\mu)/2\sigma^{2}\}\frac{(1/\sigma
^{2})^{\alpha_{0}}}{\beta_{0}^{\alpha_{0}}\Gamma(\alpha_{0})}\times\\
&  \hspace{1in}\exp\{-\beta_{0}/\sigma^{2}\}\,d(1/\sigma^{2})\\
&  =\frac{\Gamma(\alpha_{0}+1)}{\Gamma^{2}(1/2)\Gamma(\alpha_{0})}(\det
\Sigma)^{-1/2}(1+(u-\mu)^{\prime}\Sigma^{-1}(u-\mu)/2\beta_{0})^{-(\alpha
_{0}+1)}\beta_{0}^{-1}%
\end{align*}
which is the density of a $t_{2\alpha_{0}}(2,\mu,(\beta_{0}/\alpha_{0}%
)\Sigma)$ distribution.

\section{References}

\begin{enumerate}
\item Snapinn, S. M. Noninferiority trials. \textit{Current Controlled Trials
in Cardiovascular Medicine} 2000, 1, 19-21.

\item Wellek, S. Testing Statistical Hypotheses of Equivalence and
Noninferiority. \textit{Chapman \& Hall/ CRC}, 2010.

\item Gamalo, M.A., Muthukumarana, S., Ghosh, P. and Tiwari, R. C. A
generalized p-value approach for assessing noninferiority in a three-arm
trial. \textit{Statistical Methods in Medical Research} 2013, 22, 261-277.

\item Berger, R. L. and Hsu, J. C. Bioequivalence trials, intersection-union
tests and equivalence confidence sets. \textit{Statistical Science,} 1996, 4, 283-319.

\item Hauschke, D. and Hothorn, L. A. Letter to the editor. \textit{Statistics
in Medicine,} 2007, 26, 230-236.

\item Gamalo, M. A., Wu, R. and Tiwari, R. C. Bayesian approach to
noninferiority trials for proportions. \textit{Journal of Biopharmaceutical
Statistics,} 2011, 21, 902-919.

\item Gamalo, M. A., Wu, R. and Tiwari, R. C. Bayesian approach to
non-inferiority trials for normal means. \textit{Statistical Methods in
Medical Research,} 2012, online doi: 10.1177/0962280212448723.

\item Evans, M. Bayesian inference procedures derived via the concept of
relative surprise. \textit{Communications in Statistics} 1997, 26, 1125-1143.

\item Evans, M., Guttman, I., and Swartz, T. Optimality and computations for
relative surprise inferences. \textit{Canadian Journal of Statistics} 2006,
34, 113-129.

\item Baskurt, Z . and Evans, M. Hypothesis assessment and inequalities for
Bayes factors and relative belief ratios. \textit{Bayesian Analysis} 2013, 8,
3, 569-590.

\item Jeffreys, H. Some tests of significance, treated by the theory of
probability. \textit{Proceedings of the Cambridge Philosophy Society,} 1935,
31, 203-222.

\item Jeffreys, H. Theory of Probability (3rd ed.). \textit{Oxford University
Press} 1961.

\item Kass, R. E. and Raftery, A. E. Bayes factors. \textit{Journal of the
American Statistical Association,} 1995 90: 773-795.

\item Dickey, J. M. The weighted likelihood ratio, linear hypotheses on normal
location parameters. \textit{Annals of Statistics,} 1971 42: 204-223.

\item Evans, M. and Moshonov, H. Checking for prior-data conflict. Bayesian
Analysis, 1, 4, 2006, 893-914.

\item Evans, M. and Jang, G. H. Invariant P-values for Model Checking. Annals
of Statistics, 2010, 38, 1, 512-525.

\item Evans, M. and Jang, G. H. Weak informativity and the information in one
prior relative to another. Statistical Science, 2011, 26, 3, 423-439.

\item Evans, M. and Jang, G. H. A limit result for the prior predictive.
Statistics and Probability Letters, 2011, 81, 1034-1038.
\end{enumerate}

\end{document}